# Identification of a Membrane-Bound Prepore Species Clarifies the Lytic Mechanism of Actinoporins

Koldo Morante[1,2,3], Augusto Bellomio[2,3], David Gil-Cartón[4], Lorena Redondo-Morata[5], Jesús Sot[2,3], Simon Scheuring[5], Mikel Valle[4], Juan Manuel González-Mañas[2], Kouhei Tsumoto[1,6,*], and Jose M.M. Caaveiro[1,*]

*[1]Department of Bioengineering, Graduate School of Engineering, The University of Tokyo, Bunkyo-ku, Tokyo 113-8656; [2]Department of Biochemistry and Molecular Biology, University of the Basque Country, P. O. Box 644, 48080 Bilbao, Spain; [3]Biofisika Institute (UPV/EHU, CSIC), University of the Basque Country, P. O. Box 644, 48080 Bilbao, Spain; [4]Structural Biology Unit, Center for Cooperative Research in Biosciences, CICbiogune, 48160 Derio, Spain; [5]U1006 INSERM, Aix-Marseille Université, Parc Scientifique et Technologique de Luminy, 163 avenue de Luminy, 13009 Marseille, France; [6]Institute of Medical Science, The University of Tokyo, Minato-ku, 108-8639 Tokyo, Japan.*

**Running title**: *Identification of a prepore species in actinoporins*

**\*Corresponding author**: Kouhei Tsumoto (tsumoto@bioeng.t.u-tokyo.ac.jp),
and Jose M.M. Caaveiro (jose@bioeng.u-tokyo.ac.jp).

**Keywords**: pore forming protein, cytolysin, lipid- protein interaction, protein structure, oligomerization, atomic force microscopy, lipid vesicle.

## Abstract

Pore-forming toxins (PFT) are cytolytic proteins belonging to the molecular warfare apparatus of living organisms. The assembly of the functional transmembrane pore requires several intermediate steps ranging from a water-soluble monomeric species to the multimeric ensemble inserted in the cell membrane. The non-lytic oligomeric intermediate known as prepore plays an essential role in the mechanism of insertion of the class of β-PFT. However, in the class of α-PFT like the actinoporins produced by sea anemones, evidence of membrane-bound prepores is still lacking. We have employed single-particle cryo-electron microscopy (cryo-EM) and atomic force microscopy (AFM) to identify, for the first time, a prepore species of the actinoporin fragaceatoxin C (FraC) bound to lipid vesicles. The size of the prepore coincides that of the functional pore, except for the transmembrane region, which is absent in the prepore. Biochemical assays indicated that, in the prepore species, the N-terminus is not inserted in the bilayer but exposed to the aqueous solution. Our study reveals the structure of the prepore complex in actinoporins, and highlights the role of structural intermediates for the formation of cytolytic pores by an α-PFT.

Pore-forming toxins (PFT) are proteins designed for defense and attack purposes found throughout the eukaryote and prokaryote kingdoms (1,2). These proteins function by opening pores across the cell membranes, triggering processes conducive to cell death (3). PFT are commonly classified into α- and β-types according to the secondary structure of the transmembrane portion of the pore (4-6). The classical route of pore-formation begins with the interaction of the water-soluble monomer with the outer leaflet of the cell membrane, followed by the in-plane oligomerization of toxin subunits and conformational changes to assemble the lytic transmembrane pore (7). Among the intermediate species that populate this pathway, prepore particles bound to lipid bilayers have been described in the class of β-PFT (8-12). In the class of α-PFT, prepore structures of Cytolysin A from *Escherichia coli* have been proposed (6,13,14), although direct visualization still remains elusive. For actinoporins, a class of α-PFTs secreted by sea anemones, it is unclear whether a stable prepore is assembled before formation of the lytic transmembrane pore. Structurally, actinoporins are composed of a rigid β-sandwich core (mediating binding to the membrane) and two flanking α-helices, as shown for the actinoporin FraC (20 kDa, 179 residues). The N-terminal region of FraC is made of an amphipathic α-helix and neighboring residues that insert collectively in the bilayer and, together with structural lipids from the membrane, line the lytic pore (15). This remarkable metamorphosis is fully reversible under certain environmental conditions (16).

A variety of membrane-bound species have been proposed in the mechanism of actinoporins (17-19). However, the nature of some of the intermediate species and the order at which they appear during pore-formation remains unclear. Some authors have proposed that the protein subunits should first assemble into an oligomeric prepore, followed by the concerted insertion of the N-terminal region in the lipid bilayer that gives rise to the functional pore (20). Other authors, on the contrary, have suggested that the α-helix inserts deeply in the membrane prior to the oligomerization step and, therefore, this model does not contemplate the appearance of stable prepores (Figure 1) (21).

Here, we have visualized a non-lytic oligomer of FraC bound to large unilamellar vesicles (LUVs) by using cryo-EM, and to supported planar bilayers by AFM. The overall dimensions of the cryo-EM model and the high-resolution AFM images indicates that the prepore is made of eight protein subunits, a result consistent with the oligomerization number of the active pore. Biochemical assays indicate that, in the prepore species, the first few residues of the N-terminal region are not embedded in the lipid phase, but exposed to the aqueous environment. Our results reinforce the idea that protein oligomerization occurs prior to the complete insertion of the N-terminal region into the membrane, thus clarifying a critical aspect of the lytic mechanism of actinoporins.

## MATERIALS AND METHODS

*Materials.* Sphingomyelin (SM) from porcine brain and chicken egg, 1,2-dilauroyl-*sn*- glycero-3-phosphocholine (DLPC), 1,2-dipalmi- toyl-*sn*-glycero-3-phosphocholine (DPPC), and 1,2-dioleoyl-*sn*-glycero-3-phosphocholine (DOPC) were from Avanti Polar Lipids (AL, USA). 8-Aminonaphthalene-1,3,6-trisulfonic acid (ANTS), *p*-xylene-bis-pyridinium bromide, 1,1'-dioctadecyl-3,3,3',3'-tetramethylindocarbocya-nine perchlorate (DiIC$_{18}$), and Alexa Fluor 633 succinimidyl ester were from Thermo Fisher Scientific (MA, USA). Proteinase K (PK) was purchased from Sigma Aldrich (St. Louis, MO, USA).

**Protein expression and purification.** Expression and purification of FraC was carried out as previously described (22). Briefly, FraC expression was induced in *E. coli* BL21 (DE3) cells and then purified to homogeneity by ion-exchange and size-exclusion chromatography. Oxidation of the double-cysteine mutein was carried out as described by Hong et al. (23).

**Protein labeling.** To visualize FraC by fluorescence microscopy, the protein was labeled with the amine-reactive fluorescent dye Alexa Fluor 633 succinimidyl ester. The succinimidyl ester moiety of the reagent reacts with non-protonated aliphatic primary amine groups of the protein. The protein-dye conjugate was prepared following the instructions supplied by the manufacturer. Briefly, 740 µl of FraC at 180 µM in 90 mM bicarbonate buffer (pH 8.3) were mixed with 45 µl of the fluorescent dye previously dissolved in DMSO. The mixture was incubated for 1 hour at room temperature with constant stirring. To stop the reaction and remove weakly bound probes from the unstable conjugates 80 µl of freshly prepared 1.5 M hydroxylamine was added and incubated for 1 hour at room temperature. Unreacted labeling reagent was separated from the conjugate by elution of the reaction mix through a Sephadex G-15 (GE Healthcare) packed column. The protein-conjugate was tested for activity using surface pressure measurements and hemolysis assays, showing a similar behavior to that of the unlabeled protein (data not shown), ruling out detrimental effects by the dye.

**Liposome preparation and leakage assays.** LUVs of 100 nm were formed by extrusion as described previously (24). The lipid concentration was determined according to Bartlett (25). For the leakage assays, four different populations of LUVs made of DLPC, DPPC, DOPC, and SM/DOPC (1:1) were prepared as described above in a buffer containing 10 mM HEPES (pH 7.5), 50 mM NaCl, 25 mM ANTS (the fluorescent probe), and 90 mM p-xylene-bis-pyridinium (the quencher), followed by washing the liposomes with isosmotic buffer 10 mM HEPES, 200 mM NaCl, pH 7.5 in a PD-10 column (GE Healthcare). Leakage of encapsulated solutes was assayed as described by Ellens et al. (26). Briefly, LUVs were incubated with FraC at room temperature for 30 minutes to ensure, as much as possible, completion of vesicle lysis at each protein concentration employed. Upon solute release to the external medium, the dilution of quencher and fluorophore results in an increase of the emission of fluorescence of ANTS. The fluorescence was measured in a PHERAstar Plus microplate reader (BMG LABTECH, Ortenberg, Germany) with excitation/emission wavelengths of 350/520 nm. Complete release of the ANTS was achieved by solubilization of the liposomes with Triton X-100 (0.1% w/v). The percentage of leakage was calculated as:

*% leakage = $(F_f - F_0 / F_{100} - F_0) \times 100$* where $F_f$ is the fluorescence measured after addition of the toxin, $F_0$ the initial fluorescence of the liposome suspension and $F_{100}$ the fluorescence after addition of detergent.

**Surface pressure measurements.** Surface pressure measurements on lipid monolayers made of pure

DLPC, DPPC, or DOPC were carried out using a Micro-Trough-S instrument (Kibron, Finland) at room temperature with constant stirring. In these experiments, the lipid was spread over the air-water interface to the desired initial surface pressure. The protein (1 μM) was injected into the aqueous subphase and the increase of surface pressure recorded. The maximum surface pressure ($\pi_{max}$) was determined with the following equation: $\pi_{max} = \pi_0 + (\Delta\pi \cdot {''}x \, / \, b + x)$ where $\pi_0$ is the initial surface pressure, $\Delta\pi$ is the change in surface pressure, $x$ is time, and $b$ is the time necessary to reach $\Delta\pi/2$ (27). The critical pressure ($\pi_c$) corresponded to the initial surface pressure of the lipid monolayer at which the protein no longer penetrates the surface, calculated by least squares fitting as the intercept when $\Delta\pi = 0$.

***Cryo-EM of FraC Inserted in Model Membranes.***
For cryo-EM imaging, LUVs composed of DOPC were incubated with FraC (5 μM) at a protein/lipid ratio of 1:160 for 30 minutes. Holey-carbon grids were prepared following standard procedures and observed in a JEM-2200FS/CR transmission electron microscope (JEOL Europe, Croissy-sur-Seine, France) operated at 200 kV at liquid nitrogen temperature. A set of 1,562 individual pore particles were manually selected and recorded on CCD camera under low-dose conditions at 60,000 × magnification resulting in a final pixel size of 1.72 Å. An in-column omega energy filter was used to improve the signal to noise ratio of the images.

The images were CTF-corrected by flipping phases after estimation of CTF parameters in EMAN (28). The 2D images were classified by maximum-likelihood and hierarchical clustering procedures within the XMIPP software package (29). The starting 3D model was generated using reference-free alignment, classification, and common-lines procedures implemented in EMAN. This was followed by iterative refinement using a projection matching scheme in SPIDER package (30).

The rigid-body fitting was performed by maximization of the sum of map values at atom positions and by improvement in the coefficient of correlation between simulated maps from the atomic structures and the cryo-EM density map in Chimera (31). The correlation between the atomic structure of FraC and cryo-EM map suggested a resolution of ~30 Å.

***Preparation of giant unilamellar vesicles (GUVs) for confocal fluorescence microscopy.*** GUVs made of DOPC/DPPC (20:80) were prepared by electroswelling on a pair of platinum wires by a method first developed by Angelova and Dimitrov (32), modified as described previously (33). A temperature-controlled chamber was used following previous methodology (33). Briefly, a mixture of 0.2 μg/μl lipid and 0.2 % DiIC$_{18}$ was spread on the chamber and dried under vacuum. The sample was then covered with 10 mM HEPES, 200 mM NaCl, pH 7.5 at 61 ℃. This temperature was selected to prevent lipid demixing. To form the vesicles, current was applied in three steps under AC field conditions and a sinusoidal wave function: (i) 500 Hz, 0.22 V (35 V/m) for 6 minutes, (ii) 500 Hz, 1.9 V (313 V/m) for 20 minutes, (iii) 500 Hz, 5.3 V (870 V/m) for 90 minutes. After vesicle formation the chamber was left to settle at room temperature.

***Inverted confocal fluorescence microscopy.*** For the visualization of GUVs and labeled protein the chamber for GUV formation was placed on a D-Ellipse C1 inverted confocal fluorescence microscope (Nikon, Melville, NY, USA) and the samples visualized at room temperature. The excitation wavelengths used were 561 nm (for $DiIC_{18}$) and 633 nm (for Alexa Fluor 633). The fluorescence signal was collected into two different channels with band pass filters of 593/40 nm and 650 nm long pass. The objective used was a $60 \times$ oil immersion with a NA of 1.45. Image treatment was performed with the EZ-C1 3.20 FreeViewer software.

***Preparation of GUVs for AFM.*** GUVs made of SM/DOPC (1:1) were prepared by the electroswelling technique (34). A volume of 30 µl of 1 mg/mL lipids dissolved in chloroform: methanol (3:1) were deposited in two glass plates coated with indium tin oxide (70-100 Ω resistivity, Sigma-Aldrich) and placed in the desiccator at least 120 minutes for complete solvent evaporation. A U-shape rubber piece of ~1 mm thickness was sandwiched between the two indium tin oxide side slides. Then the formed chamber was filled with ca. 400 µl of 200 mM sucrose and exposed to 1.2 V AC current (12 Hz sinusoidal for 2h, 5 Hz squared for 10 minutes). The resulting suspension was collected in a vial and used within several days. ***Supported lipid bilayer preparation for AFM.*** A total of 1 µL of a suspension of GUVs was deposited onto freshly cleaved 1 mm² mica pretreated with 1 µL of 10 mM Tris-HCl, 150 mM KCl, pH 7.4 (imaging buffer) and incubated for 15 minutes at room temperature. The resulting supported lipid bilayers were carefully rinsed with imaging buffer before image collection and always kept under aqueous environment. During imaging, FraC toxin was injected into the fluid cell to give a final concentration of ~10 µM.

***AFM imaging.*** AFM was performed at room temperature on a high-speed AFM 1.0 instrument (RIBM, Japan) equipped with short high-speed AFM cantilevers (~8 µm, NanoWorld, Switzerland) with nominal resonance frequency of ~1.2 MHz and ~0.7 MHz in air and liquid, respectively, and a nominal spring constant of ~0.15 Nm⁻¹. Image acquisition was operated using optimized feedback by a dynamic PID controller. Small oscillation free ($A_{free}$) and set point ($A_{set}$) amplitudes of about 1 nm and 0.9 nm, respectively, were employed to achieve minimum tip-sample interaction. Typically, pixel sampling ranges from $100 \times 100$ pixels and $200 \times 200$ pixels and frame rate between 500 and 800 ms per frame.

***AFM data analysis.*** AFM data was analyzed in ImageJ and with self-written image analysis scripts (movie acquisition piezo drift correction) in ImageJ (35). To obtain the high resolution images shown in Figure 5 and Figure 6, five consecutive frames were time-averaged. All further analysis, i.e. histogram distributions were analyzed in Igor and Origin.

***Protease susceptibility assay.*** Proteinase K (PK) (50 µM) was incubated with FraC (50 µM) for 24 hours at room temperature in 50 mM Tris, 200 mM NaCl, 5 mM $CaCl_2$ at pH 7.4. In the assays with lipids, FraC was incubated with the appropriate LUVs (7.5 mM) made of either DOPC or SM/DOPC (1:1) for 30 minutes prior to the addition of PK. The reaction was stopped by adding phenylmethylsulfonyl fluoride at a final concentration of 5 mM for 10 minutes and analyzed by SDS-PAGE.

***N-terminal sequencing.*** SDS-PAGE protein bands of the PK reaction products were transferred to a polyvinylidene fluoride membrane (Bio-Rad Laboratories, CA, USA), stained with Ponceau 3R solution for 1 hour, and washed with water until complete removal of the excess stain. The red colored protein bands were excised from the membrane and their N-terminal sequenced employing standard techniques (36).

## RESULTS

### Interaction of FraC with PC membranes –

Because actinoporins are specifically activated by membranes containing the lipid SM, the use of lipid compositions in which SM is absent but where the toxin maintain a strong interaction with the vesicles may reveal structural intermediates not detectable by other means. We first evaluated the interaction of FraC with vesicles made of various types of the lipid phosphatidylcholine (PC, a lipid displaying the same phophocholine headgroup moiety as that of SM) to determine the optimum PC species yielding the highest possible association between protein and liposomes. The three PC lipid species examined were DLPC, DPPC, and DOPC each differing in the length and degree of saturation of their acyl chains. To evaluate the degree of interaction of FraC with these lipids we measured the magnitude of the insertion of the protein in a monolayer of lipid molecules at the water-lipid interface (Figure 2A). We determined the surface pressure at which the protein will no longer penetrate, known as critical pressure ($\pi_c$) (27). The lipid composition at which $\pi_c$ was highest corresponded to that of monolayers composed of DOPC ($\pi_c$ = 36.1 ± 1.6 mN/m) followed by that of monolayers made of DLPC ($\pi_c$ = 31.1 ± 1.3 mN/m). The insertion of FraC in DPPC monolayers was meager (23.6 ± 1.0 mN/m). These data suggest that the toxin associates more readily with lipids in the liquid-expanded phase such as DOPC and DLPC than those in the liquid-condensed phase (DPPC) (37-39). However, the protein does not generate pores in LUVs when SM was absent regardless of the lipid phase (Figure 2B). A close association between actinoporins and lipid monolayers thus does not guarantee effective formation of pores in a lipid bilayer system as there are other physicochemical properties involved in pore formation, such as lipid phase coexistence and the presence of SM (40).

Additional evidence describing the lipid preference of FraC was gathered by visualizing the binding of the fluorescently-labeled toxin to GUVs composed of DOPC/DPPC (20:80). In these experiments, the fluorescent dye conjugated to the toxin co-localized with the DOPC domains (dark areas in Figure 2C) indicating that the toxin preferentially binds to the fluid phase domains over the gel domains, a result consistent with the observations made with monolayers for FraC and leakage assays for sticholysin II (41). Based on these results, membranes composed of DOPC were selected for structural studies analyzing the conformation of FraC bound to membranes in a non-lytic environment.

***Cryo-EM*** − To visualize the structure of membrane-bound FraC, vitrified samples of toxin-treated DOPC liposomes were imaged by cryo-EM. The observed ring-shaped particles covering the lipid vesicles were attributed to protein oligomers (Figure 3A). A total of 1,562 top- and side-view images were selected to build a three-dimensional model of the protein oligomer. The model was built by common-lines procedures, followed by iterative refinement using projection matching of the class-averaged images and the density map projections (Figure 3B). A second classification method based on maximum-likelihood (42) and hierarchical clustering approaches (43) (Figure 3C) rendered images similar to those used to generate the final density map.

 The reconstructed image consisted of a doughnut-shaped ring with an external and internal diameter of ~11 and ~ 5 nm, respectively (Figure 3D and 3E). These dimensions are very similar to those of the crystallized pore in the active state (15). However, unlike the cryo-EM model of the pore bound to SM/DOPC (1:1) liposomes (20), the oligomer bound to DOPC vesicles does not span the lipid membrane (see below). This architecture is consistent with a non-lytic oligomeric species resembling a prepore. A rigid-body fitting of an octameric model of FraC based on the atomic structure of the transmembrane pore of FraC achieved a high cross-correlation coefficient (cc = 0.82, Figure 3D, E) (15). The oligomeric model fits well within the perimeter of the cryo-EM map, except for the N-terminal region, which lies outside the electron density map suggesting that it is either resting on the surface of the membrane or inserted in the hydrophobic core of the membrane (21,23,44). A nonamer of FraC mimicking the structure of a crystallized oligomer of FraC in the presence of detergents (20,45) was also fitted in the cryo-EM maps, yielding a cross-correlation coefficient only slightly worse (cc = 0.81) than that of the octamer. The fitted nonamer displayed a few clashes between protomers, in contrast to an oligomer made of ten units in which the numerous collisions between protein chains made the decamer prepore unfit for this electron density. From these data we cannot rule out the existence of a minor population of nonameric prepore species preceding a hypothetical nonameric pore, as discussed previously (15).

 A comparative analysis of the electron density distribution along the central section of the oligomer of FraC in DOPC membranes and in SM/DOPC (1:1) clearly shows that the differences between pores and prepores occur in the critical transmembrane region. To perform this comparison we employed the previously reported cryo-EM map of the active pore (20). For the analysis, 3D volumes focused on the pore regions (shown within yellow rectangles in Figures 4A and 4B) were projected into 2D images, and the gray values of the images (resulting from the accumulation of 3D density values) were plotted in 1D profiles (Figures 4C and 4D). The structure of FraC in DOPC membranes (Figure 4A) reveals a peak of higher density values in a central lobe at the membrane level below the vestibule of the oligomer (Figure 4C), whereas in SM/DOPC (1:1) membranes (Figure 4B), the same region is characterized by lower density values (Figure 4D). These features are consistent with the absence (in DOPC membranes) or the presence (in SM/DOPC (1:1) membranes) of a transmembrane pore. In the former, the high-density central region likely represents the accumulation of membrane lipids and N-terminal α-helices detached from the β-core of the toxin. In contrast, in membranes containing SM the central region of the oligomer displays lower electron density because the α-helices span the membrane from top to bottom, and consequently the lipids are cleared off producing an aqueous pore.

**AFM** − In the presence of supported lipid bilayers composed of the equimolar mixture SM/DOPC (1:1), WT FraC assembles in a dense array of closely packed oligomers as determined by AFM (Figure 5). These oligomers, presumably corresponding to pore particles, cover the SM-rich domains in an arrangement previously observed in FraC and other actinoporins (45,46) or the SM-specific PFT lysenin (47). The cross-section profile of the oligomeric complexes reveals an average diameter of $7.5 \pm 0.6$ nm, a value in good agreement with the mean diameter (average of outer and inner diameter) of the pore determined by X-ray crystallography (~ 8 nm). Eight protein chains are observed in three well-resolved pore particles encountered (see for example Figure 5C). Because prepores of FraC were not resolved in DOPC, probably caused by high-diffusivity preventing AFM contouring, a construct of FraC bearing a double cysteine mutation (V8C/K69C, termed 8-69[OX]) was instead examined on supported membranes made of SM/DOPC (1:1).

Under oxidizing conditions, the N-terminal segment of this mutein is covalently attached to the protein core by means of a disulfide bond, preventing the protein from generating a transmembrane pore, and thus inactivating the toxin (15,23). As with WT FraC, the construct 8-69[OX] also gave rise to a dense array of pore-like particles (Figure 6), indicating that the protein readily oligomerizes in the presence of membranes even if the N-terminal region remains attached to the protein. The average diameter of these particles ($6.2 \pm 0.7$ nm) is somehow smaller than that of WT protein, reflecting the influence of the N-terminal region attached to the β-core region. Because of the constrains imposed by the disulfide bond, the conformation of the N-terminal region in 8-69[OX] is likely to differ from that of WT FraC bound to liposomes made of DOPC (Figure 3D, E). To further investigate this question we employed biochemical assays (see below).

***Protease susceptibility of the membrane-bound toxin*** − It was shown that FraC bound to LUVs exhibits different susceptibility to proteinase K (PK) depending on the lipid composition of the membrane (15). The incubation of FraC with PK generated a product of smaller size when the toxin was bound to DOPC vesicles as compared to those generated in the presence of SM/DOPC (1:1) vesicles (15), although the basis of this difference was not explained. In view of the new prepore oligomeric species described herein, we hypothesized that the N-terminus of this prepore is located in a solvent-exposed environment accessible to PK, whereas in the pore the N-terminus is deeply inserted in the membrane and thus inaccessible to the protease. To verify this hypothesis and determine the extent of the digestion, we incubated samples of FraC with PK followed by their separation by SDS-PAGE and N-terminal sequencing.

The incubation of PK with FraC in the presence of DOPC vesicles yields a fragment of smaller molecular weight than that of the untreated protein (shown in the 20 kDa region). In contrast, in the presence of SM/DOPC (1:1) vesicles, the bands of treated and untreated toxin display the same molecular mass (Figure 7A). The mutein 8-69[OX] bound to membranes was also employed, since its N-terminus remains exposed to the solvent constrained by the disulfide bond. As expected, 8-69[OX] was also susceptible the proteolytic activity of PK in DOPC and SM/DOPC (1:1) vesicles. To determine the cleavage point the proteins were sequenced from their N-terminus. The sequencing data revealed that, in the presence of vesicles of DOPC, FraC WT and 8-69[OX] were cleaved at the N-terminus by PK, rendering products in which the first four and first eleven residues, respectively, were missing (Figure 7B, C). FraC bound to SM/DOPC (1:1) was not digested by PK as expected from the position of the band in the SDS-PAGE gel, whereas 8-69[OX] was cleaved at the same position seen in vesicles of DOPC. These results demonstrate that the N-terminus of FraC in DOPC vesicles (prepore configuration) is accessible to PK, i.e. this region is not embedded in the lipid bilayer.

## DISCUSSION

Structural intermediates that populate the pathway leading to the formation of a functional transmembrane pore in PFT are key species that can help to elucidate the details of pore formation. Membrane-bound oligomeric structures poised for membrane disruption are commonly referred to as prepores and have been visualized in lipid bilayers only for β-PFT. In contrast, the existence of prepores in α-PFT is controversial. An example is the family of actinoporins, where a strong debate is held about the existence or not of these non-lytic oligomers (20,21,48). Until now, the evidence supporting a prepore in actinoporins was based on the crystal structure of a non-lytic nonameric ensemble solved for FraC (20).

Herein, we have described a low-resolution membrane-bound oligomer consistent with the ability of FraC to assemble as a prepore on biological membranes. We employed a protein concentration above physiological levels to ensure a large and homogeneous population of pre-pore species bound to the liposomes, thus facilitating their visualization by cryo-EM and AFM. The pre-pore structure could explain the readiness of actinoporins to induce lysis in liposomes made of PC upon generation of lipid domains *in situ* (40). The size and stoichiometry of the prepore in the cryo-EM (Figures 3, 4) and AFM (Figure 6) images were in the range of those of the crystallized pore species (15). The cryo-EM reconstruction map is not consistent with a transmembrane pore, an argument strengthened by the comparison side-by-side with cryo-EM data of pores of FraC embedded in vesicles of SM/DOPC (1:1) (20).

Electron density gradient analysis and protease digestion assays suggest a close association of the N-terminal region with the membranes in a position approximately parallel to the plane of the membrane as was described before for other actinoporins (44,49). Evidence that this oligomer precedes pore formation is inferred from a previous study carried out with the actinoporin equinatoxin II. It was shown that the addition of phospholipase C to vesicles of PC decorated with toxin promoted vesicle lysis by the *in situ* generation of lipid domains (40). Our results suggest a model where the N-terminal α-helices penetrate the bilayer in a concerted manner (Figure 8), an alternative mechanism to that in which helix penetration occurs before protein oligomerization (21). Although pore-formation by the successive insertion of single α-helices cannot be completely ruled out in membranes made of SM/DOPC (1:1), simple thermodynamic considerations suggest that would not be the case: The penetration of individual α-helices containing a large number of charged residues (FraC displays three Asp and one Glu in

### ACKNOWLEDGEMENTS

this region) in the hydrophobic core of biological membranes would be strongly disfavored (50-52). In conclusion, our study clarifies the structure of a key intermediate, known as prepore, in the route of pore formation by actinoporins belonging to the group of α-PFTs. The characterization of the prepore in actinoporins highlights similarities with the mechanism for pore formation of the group of β-PFT, despite these two groups having quite distinct architectures at the transmembrane region.

AB is a staff scientist from the CONICET (Argentina) and received a visiting scientist fellowship from the Basque Government while conducting this work. We thank Dr. S. Kudo for expert advice. This work was supported by a Grant-in-Aid for Scientific Research A (25249115 to KT) and a Grant-in-Aid for Scientific Research C (15K06962 to JMMC). KM was a recipient of a fellowship from the Spanish Ministerio de Ciencia e Innovación during the beginning of this study. Work in the Scheuring-Lab was supported by a European Research Council Grant (#310080, MEM-STRUCT-AFM). This work was also supported by grant BFU2015-66326-P from the Spanish Ministry of Economy and Competitiveness to MV.

### FOOTNOTE

The present address of AB is Instituto Superior de Investigaciones Biológicas (INSIBIO, CONICET-UNT) e Instituto de Química Biológica "Dr. Bernabé Bloj", Facultad de Bioquímica, Química y Farmacia, Universidad Nacional de Tucumán, Chacabuco 461, San Miguel de Tucumán. Argentina.

### CONFLICT OF INTEREST

The authors declare that they have no conflicts of interest with the contents of this article.

### AUTHOR CONTRIBUTIONS

KM, JMGM, and JMMC conceived the study. KM, JMGM, KT and JMMC coordinated the study. MV and DGC analyzed and performed cryo-EM studies. SS and LRM analyzed and performed AFM studies. KM,

AB, DGC, LRM, and JS performed experiments. All authors designed experiments and analyzed the data. KM and JMMC wrote the paper with input from all other authors. All authors reviewed the results and approved the manuscript.

## ABBREVIATIONS

PFT, pore-forming toxins; FraC, fragaceatoxin C; SM, sphingomyelin; DLPC, 1,2-dilauroyl-*sn*-glycero-3-phosphocholine; DPPC, 1,2-dipalmitoyl-*sn*-glycero-3-phosphocholine; DOPC, 1,2-dioleoyl-*sn*-glycero-3-phosphocholine; DiIC18, 1,1'-dioctadecyl-3,3,3',3'-tetramethylindocarbocyanine perchlorate; LUVs, large unilamellar vesicles; GUVs, giant unilamellar vesicles; AFM, atomic force microscopy; cryo-EM, single-particle cryo-electron microscopy; PK, Proteinase K; CTF, contrast transfer function.

## <u>FIGURES</u>

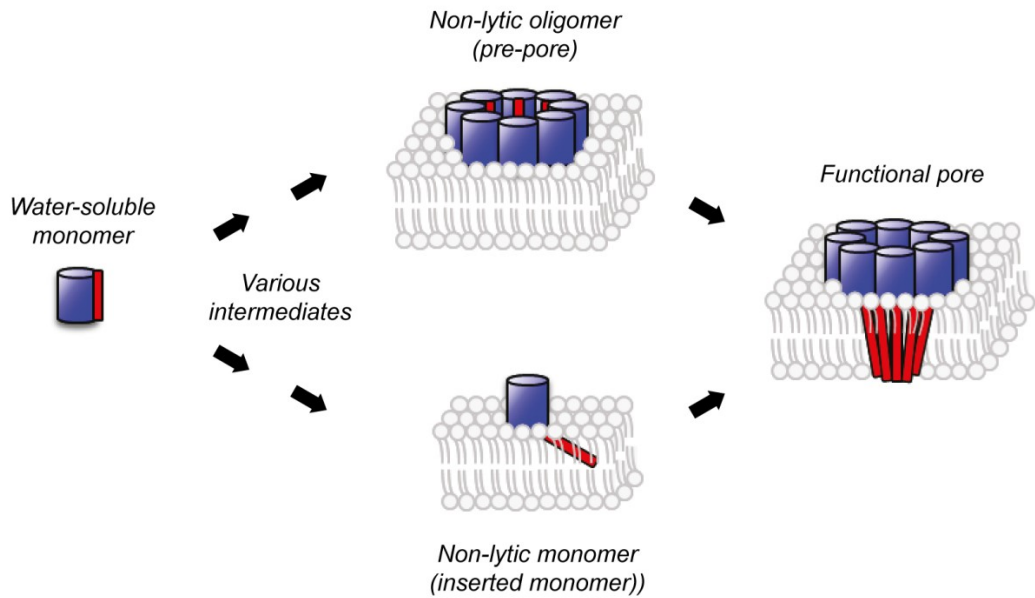

**Figure 1. Two alternative routes for pore formation in actinoporins.** The binding of the water-soluble monomer to the cell or model membranes leads to a lytic (active) pore by at least two alternative routes, as shown

in the figure. Top, formation of a non-lytic oligomer (prepore) precedes insertion into the membrane (20). Bottom, insertion of the N-terminal region into the membrane occurs prior oligomerization of the functional pore (21).

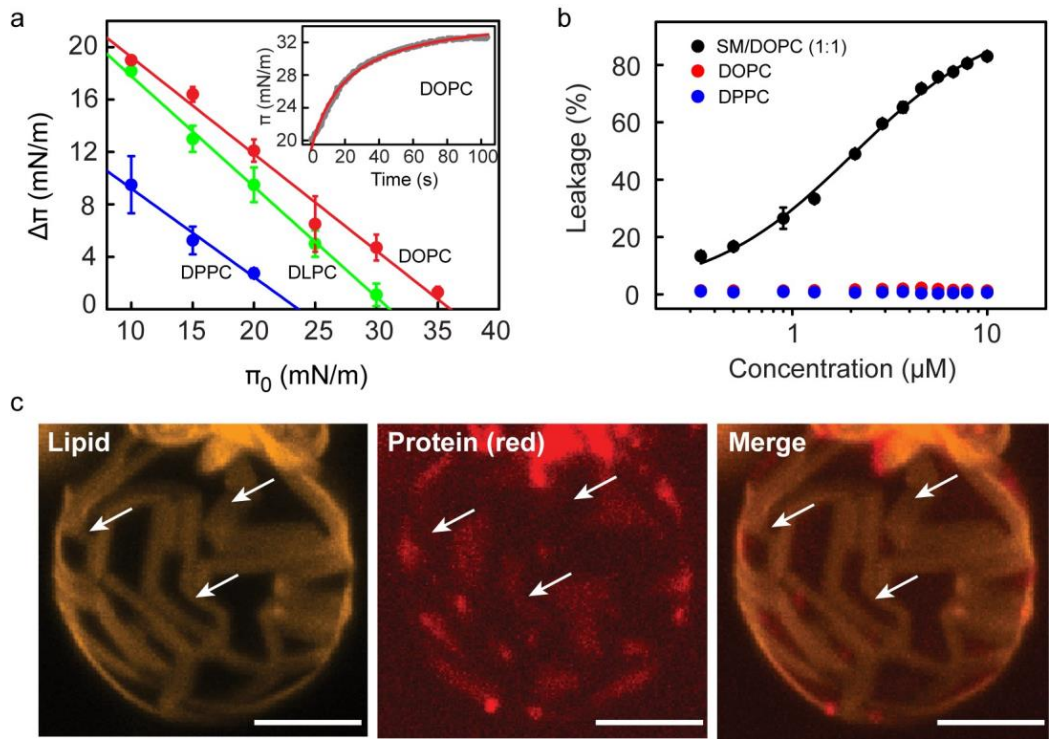

**Figure 2. Interaction of FraC with model membranes.** (**a**) Change in surface pressure of lipid monolayers composed of DOPC (red), DLPC (green), or DPPC (blue) after treatment with FraC (1 μM). The parameter $\pi_c$ corresponds to the value of $\pi_0$ where the regression line intersects the abscissa. The inset shows a representative example of the kinetic profile of insertion of FraC in DOPC monolayers ($\pi_0 = 20$ mN/m, gray trace). The experimental data was fitted to a hyperbola (red line) from which the value of $\Delta\pi$ was determined. (**b**) Lytic activity of FraC in LUVs made of DOPC (red) or DPPC (blue). The data obtained with SM/DOPC (1:1) represents a positive control (black). The LUVs made of DLPC are permeable to encapsulated dyes in the absence of protein (spontaneous leakage) and thus the data obtained with them was not considered. For the experiments in panels (a) and (b) the mean and standard deviation of three independent measurements was plotted. (**c**) Binding of FraC to GUVs made of DOPC/DPPC (20:80) supplemented with 0.2% DiIC$_{18}$. This probe partitions in the ordered phase regions (yellow domains) (53,54). Protein (red) was added to a final concentration of 1.3 μM. Lipid and protein were visualized with a 593/40 nm band pass filter (yellow, left panel), or with a 650 nm long pass filter (center panel), respectively. Merged images are shown on the right panel. The white arrows point at liquid disordered regions (dark domains) where FraC is preferentially located.

The scale bar represents 5 μm.

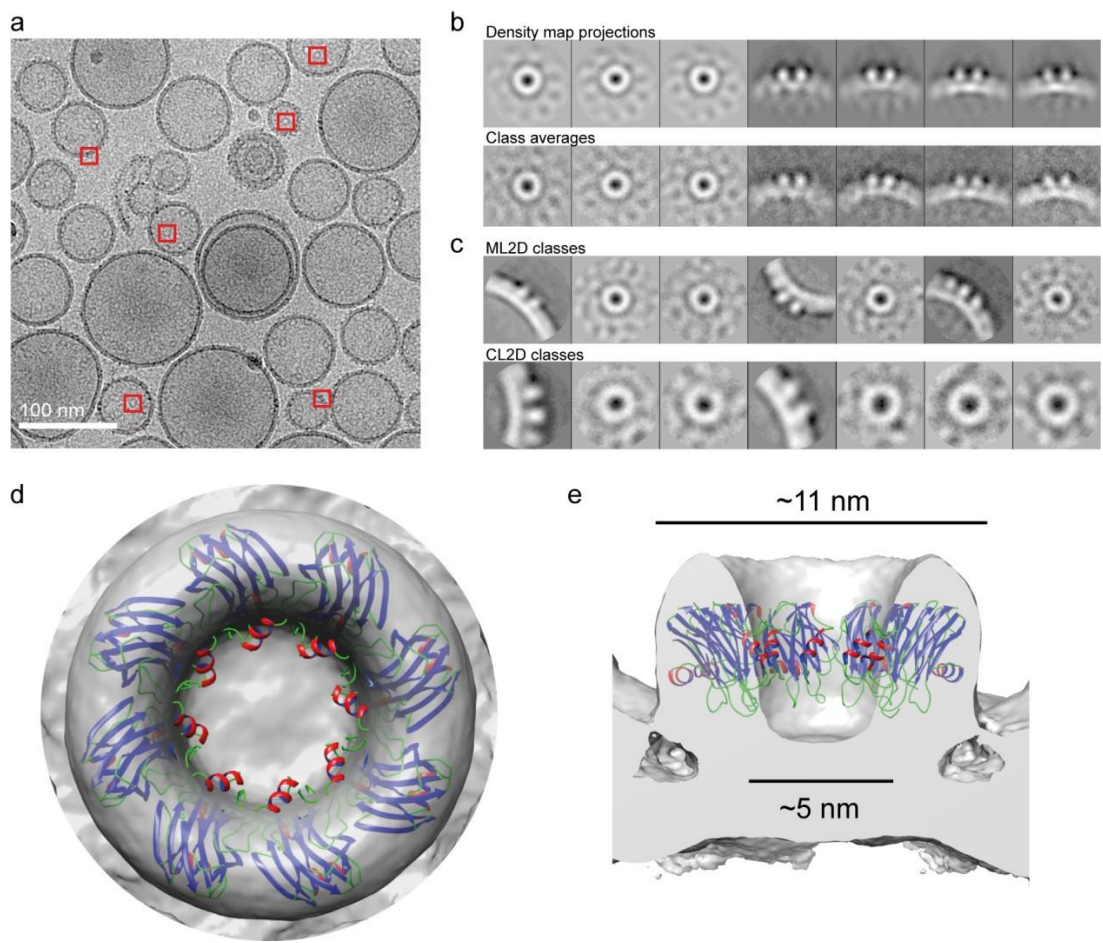

**Figure 3. Structure of the oligomeric prepore of FraC bound to DOPC vesicles.** (**a**) Representative image of FraC in DOPC vesicles obtained by cryo-EM. Top- and side-views of the protein oligomers were selected (red squares) for subsequent classification analysis. The scale bar corresponds to 100 nm. (**b**) Density map projections (top row) and 2D class-averaged particles (bottom row) employed to build a 3-dimensional model of the protein oligomer (see below). (**c**) Set of particles obtained by maximum-likelihood (ML2D, top row) and hierachical clustering (CL2D, bottom row) procedures. (**d**) Top- and (**e**) side-views of the 3-dimensional model of the prepore of FraC bound to vesicles of DOPC. The atomic model of FraC was built as an octamer using the coordinates of the protomer of FraC prior to pore formation (entry code 4TSL).

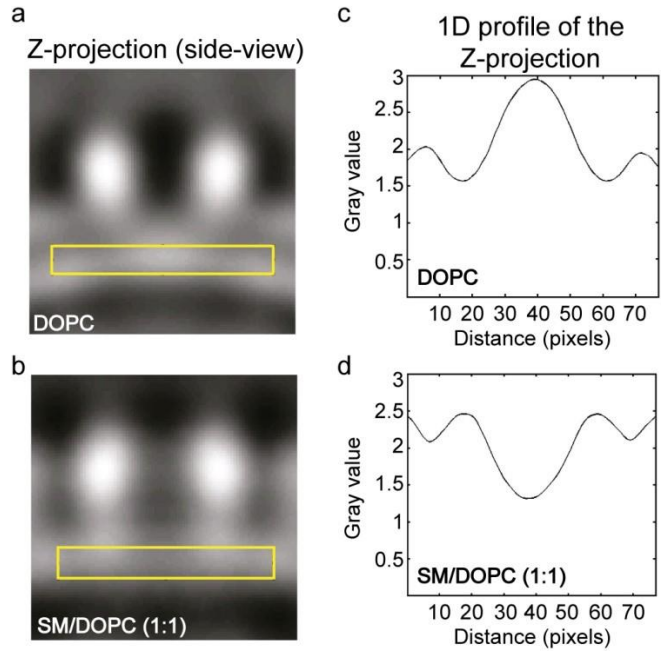

**Figure 4. Electron density of FraC bound to vesicles.** Side view (Z-projection) of oligomers of FraC bound to vesicles of (**a**) DOPC, or (**b**) SM/DOPC (1:1). The yellow square indicates the region where the 1-dimensional profile of the Z-projection (shown in **c, d**) was calculated. The intensity of the electron density is expressed in gray values. Panels (**b**) and (**d**) correspond to the analysis carried out with published data (20), although we note that the analysis presented here has not been shown elsewhere.

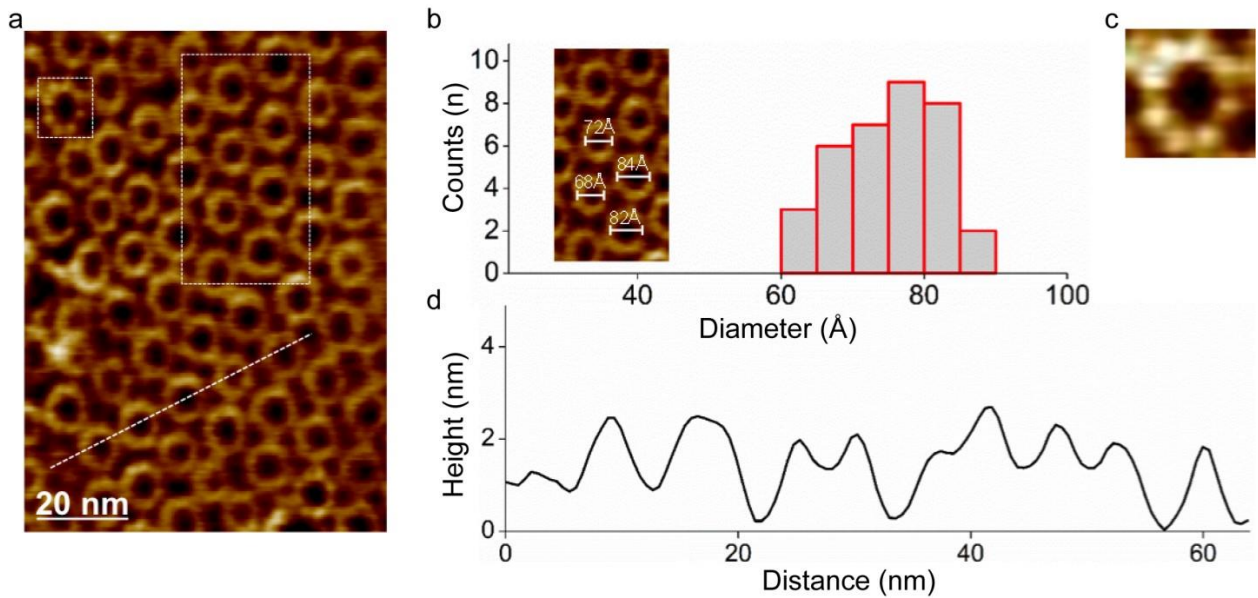

**Figure 5. Visualization of pores of WT FraC with AFM.** (**a**) Two-dimensional packing of ring-shaped oligomers of WT FraC on supported lipid bilayers composed of the lipid mixture SM/DOPC (1:1). (**b**) Diameter distribution analysis (peak-to-peak distances of the protein protrusion in the height profile). The average diameter of the particles was $75 \pm 6$ Å (mean $\pm$ SD from the Gaussian distributions). Inset: detail of the particles inside the white dashed rectangle in panel (a). (**c**) Magnification (13 nm frame size) of a single FraC oligomer in panel (a) (white dashed square). (**d**) Cross-section profile (left to right) of FraC oligomers shown in panel (a) (white dashed line). The molecules are packed with a center-to-center distance of ~112 Å.

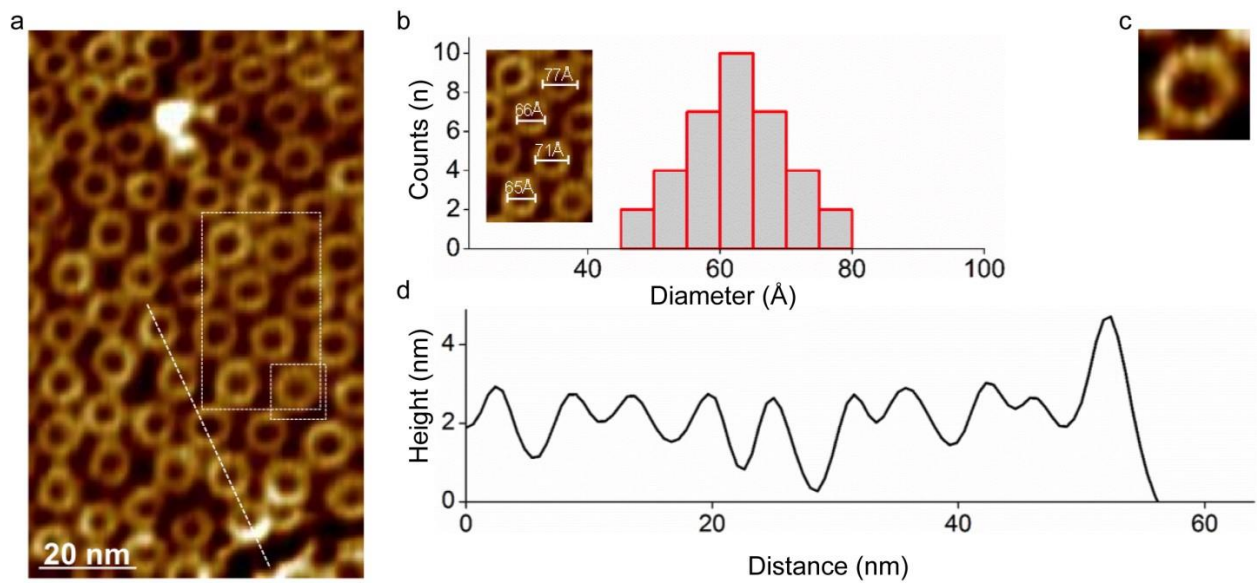

**Figure 6. Visualization of prepores of 8-69$^{OX}$ FraC with AFM.** (**a**) Two-dimensional packing of ring-shaped oligomers of 8-69$^{OX}$ FraC on supported lipid bilayers composed of the lipid mixture SM/DOPC (1:1). (**b**) Diameter distribution analysis (peak-to-peak distances of the protein protrusion in the height profile). The average diameter was $62 \pm 7$ Å (mean $\pm$ SD from the Gaussian distributions). Inset: detail of the particles inside the white dashed rectangle in panel (a). The slightly smaller diameter compared to the WT suggests a tighter association of the subunits in the 8-69$^{OX}$ FraC mutant. (**c**) Magnification (12 nm size frame) of a single prepore of FraC 8-69$^{OX}$ (white square in panel a). (**d**) Cross-section profile (left to right) of prepore particles of FraC 8-69$^{OX}$ (white line in panel a). The molecules packed with a center-to-center distance of ~108 Å.

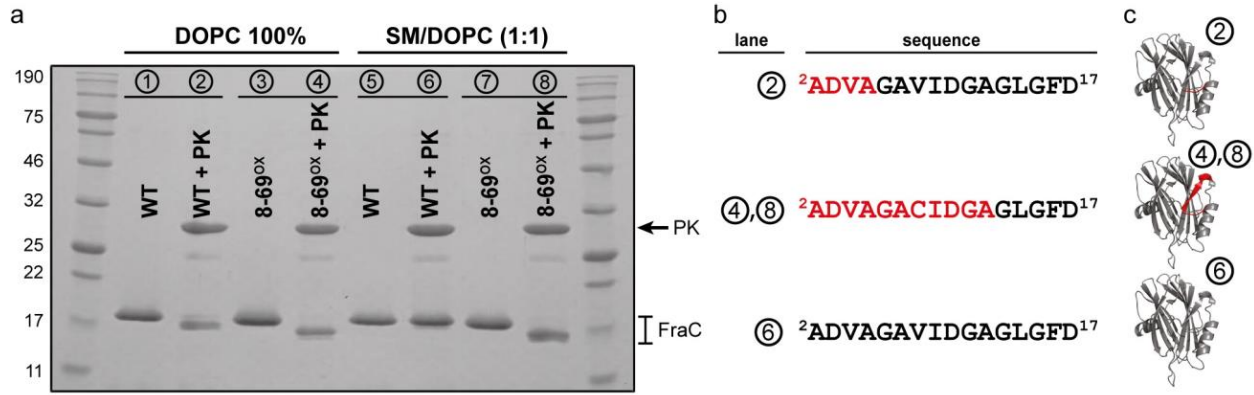

**Figure 7. Protection of FraC from PK in the presence of liposomes. (a)** SDS-PAGE of the products obtained after the incubation of WT and 8-69$^{OX}$ FraC with DOPC vesicles in the absence and in the presence of PK (lanes 1-4) or with SM/DOPC (1:1) (lanes 5-8). **(b)** N-terminal sequence of FraC after digestion with PK. The circled number before the sequence corresponds to the lane of the same number in the SDS-PAGE. Residues highlighted in red were digested by PK. The first 16 residues of the recombinant WT protein expressed in *E. coli* are ADVAGAVIDGAGLGFD (55). **(c)** The location of the residues digested by PK are depicted in the three dimensional structure of the monomer of FraC (PDB code 3VWI).

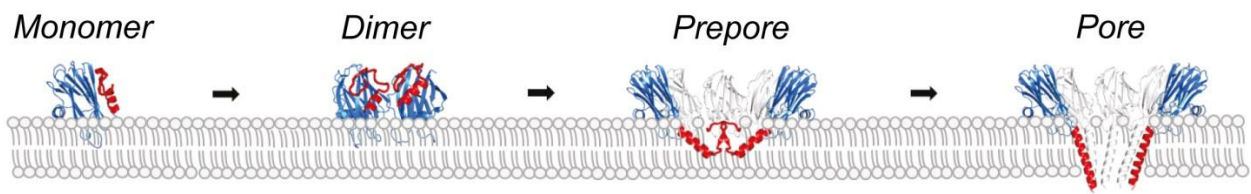

**Figure 8. Model for pore formation by FraC.** A toxin monomer binds the membrane. The membrane promotes protein-protein interactions between monomers to produce a dimer (15) leading to prepore upon successive addition of monomer and/or dimers to the growing oligomer. The N-terminal α-helices in the prepore embedded on the surface on the membrane with the N-terminus exposed to the aqueous solution. The conversion to the transmembrane pore would be achieved by the concerted penetration and elongation of the helices across the lipid bilayer. The structures of the monomer, dimer, and pore were retrieved from the PDB with entry codes of 3VWI, 4TSL, and 4TSY, respectively. The structure of the prepore at high resolutions has not been determined experimentally. In this figure the structure of the prepore is drawn to illustrate the model consistent with the experimental data reported in our study.